\documentclass[aps,pra,twocolumn,superscriptaddress,nofootinbib]{revtex4-1}
\usepackage{amssymb,graphicx,color}
\usepackage[intlimits]{amsmath}
\usepackage[english]{babel}
\usepackage[colorlinks]{hyperref}

\def\beq{\begin{equation}}
\def\eeq{\end{equation}}
\def\bea{\begin{eqnarray}}
\def\eea{\end{eqnarray}}
\def\barr{\begin{array}}
\def\earr{\end{array}}

\newcommand{\vect}[1]{\boldsymbol{#1}}

\begin{document}

\title{Survival of weak-field seekers inside a TOP trap}

\author{Nirupam Dutta}
\email[E-mail:]{nirupamdu@gmail.com}
\affiliation{Institute of Physics, Sachivalaya Marg, Sainik School PO, Gajapati Nagar, Bhubaneswar, Odisha 751005, India}

\author{Anirban Dey}
\email[E-mail:]{deyanirban89@gmail.com}

\author{Prasanta K. Panigrahi}
\email[E-mail:]{pprasanta@iiserkol.ac.in}
\affiliation{Indian Institute of Science Education and Research Kolkata, Mohanpur, West Bengal 741 246, India}

\date{\today}

\begin{abstract}

In this article, for the first time in the context of TOP trap, the necessary and sufficient conditions for the adiabatic evolution of weak field seeking states have been quantitatively examined. It has been well accepted since decades that adiabaticity has to be obeyed by the atoms for successful magnetic trapping.  However, we show, on the contrary, that atoms can also be confined  beyond the adiabatic limit. Hence, our findings open new possibilities to relax the restrictions of atom trapping in laboratories.

\end{abstract}

%\pacs{}

\maketitle

\section{Introduction}
Magnetic trapping is a commonly used technique \cite{zero, one, two, three, four, five} to confine atoms within a small region of space in order to study their precise quantum mechanical behaviour in the laboratory.  Wide range of experimental studies on ultracold atoms exploit such techniques and remarkable experimental success has been achieved in this direction in recent times. To overcome the drawbacks of magnetic trapping, various additional ideas in both technical and theoretical domains have also been introduced. One of the improved traps, which is the central interest of this article, is the Time Orbiting Potential (TOP) trap \cite{one, three, five, six}. It was successfully exploited to confine bosons at an ultracold temperature to achieve one of the noble states of matter, Bose-Einstein condensate (BEC) \cite{zero, seven, eight, nine}.

A TOP trap is a modified version of the quadrupole magnetic field trap (QMFT), where an additional rotating magnetic field is added along with the original magnetic field of QMFT.
In QMFT, the atoms which are weak field seekers \cite{nine} are attracted towards the minima of the magnetic field. On the other hand, the strong field seekers fly off the trap \cite{nine}. Furthermore, the minimum in QMFT has a zero magnetic field, which makes the separation between the Zeeman levels of the atom negligibly small (practically zero). Hence, weak field seekers in the vicinity of the minima flip to strong field seeking states and leave the trap \cite{one}. To prevent this from happening, a rotating magnetic field is added to QMFT and this new trap is called  a TOP trap. 

The rotating field component of the TOP trap forces the minimum of the magnetic field to rotate in a circle of radius $R_0=\frac{B_0}{A_0}$ \cite{four, han}, where, $B_0$ is the magnitude of the rotating field and $A_0$ designates the static magnetic field configuration of QMFT. Until now, in the analysis of TOP trap, a time averaged description has been adopted, which makes the minimum of the field configuration a non-zero time independent quantity. Therefore, by avoiding the possible flipping between weak and strong field seeking states, TOP trap can be utilised successfully to trap atoms in the laboratory. The concept of time averaged field is physically allowed when the rotation frequency $\omega$ of $B_0$ is large enough. In principle, to validate the analysis at arbitrarily small time scale, one must introduce a rotating field with a frequency $\omega$, which is infinitely large.

Besides this, TOP trap involves other important scales like Larmor frequency $\omega_0$ of the atom, which is proportional to the magnetic field strength of the trap. To avoid the possibility of spin flipping transition, the adiabatic condition $\omega << \omega_0$  \cite{three, five, eight} has to be maintained. Therefore, for a successful trapping of neutral atom, the rotation frequency $\omega$ cannot be made arbitrarily large. One can always introduce very high intensity magnetic field, such that $\omega_0$ itself becomes very large and hence, time averaged description becomes possible to describe the system even in  a very small time-scale. However, in a weak magnetic environment, the concept of time-averaging field is completely impossible in the light of the analysis presented above.\\
Until now, we have not mentioned anything about the oscillation of an atom inside the trap but that oscillation frequency is also a crucial scale for the trap under consideration. To be precise, for a successful TOP trapping, the following condition has to be obeyed \cite{six, ten, eleven, sherlock, igor}:
\begin{equation} \label{eq:1}
\omega_{osc}<<\omega<<\omega_0,
\end{equation}
where, $\omega_{osc}= \sqrt{\frac{k}{m}}$, is the frequency of oscillation of atom cloud inside the trap and $k$ is given by, $k=\frac{|\mu|A_0^2}{2B_0}$. Here, $\mu$ is the magnetic moment of the atom and $m$ is the atomic mass.  \\
In order to make $\omega_{osc}$ arbitrarily small, one must reduce the ratio $\frac{A_0}{B_0}$ arbitrarily, which on the other hand make the radius of the circle of death ($R_0$) \cite{clear, nine} arbitrarily large. Therefore, we need a judicious arrangement of these three different frequencies. Eventually the time scale of this spatial oscillation of the atoms is large compared to that of the rotating bias field and the dynamics of the spin states\cite{six, ten}. Hence, by exploiting the slow motion of the atom cloud, we can always maintain the condition $\omega$, $\omega_0 >> \omega_{osc}$ with a proper choice of TOP trap parameters. Hence, Eqn.\ref{eq:1} is a sufficient criterion for trapping of weak field seekers inside a TOP trap. Here, we would like to raise the question: {\it is it necessary to satisfy the criterion $\omega<<\omega_0$ to ensure adiabatic evolution of weak field seeking states?} The answer is no and we will see that explicitly in this article. Therefore, we can relax the condition (1) and can avoid the difficulties arising from the complex adjustment of these above mentioned frequencies (or time scales). Furthermore, {\it we would like to know whether adiabatic evolution of weak field seeking state is at all necessary for a successful TOP trapping}. Our present work is devoted towards addressing these two relevant queries. \\

We have organised the article in the following way. After the introduction, in Section II, we have discussed the necessary and sufficient criteria for adiabatic evolution of weak field seeking states at any given parametric position of atoms by exploiting the separability of the dynamics of spin state and spatial wave function. In Section III, for a given parametric position, we have examined the time evolution of the spin state by solving Schr\"odinger equation for a time varying Hamiltonian without adopting any time averaged description of the magnetic field. Finally, in section IV we have evaluated the spatial average of survival probability for weak field seeking state by considering the motion of the atom cloud along with evolution of spin degrees of freedom. At the end, we have summarised our investigations and have discussed the importance and possibilities of our findings in the domain of cold atom physics.

\section{Criterion for adiabatic evolution of weak field seekers} 
To study the evolution of the  atom cloud, we consider them initially in  the weak field seeking state. As the spatial motion of the weak field seekers can be separated from the spin dynamics, we can always treat the spatial degrees of freedom parametrically. Therefore, one first needs to investigate the dynamics of spin part of the state for a fixed space point and then the dynamics of spatial degrees of freedom should be appropriately taken care off. This is very  similar to the Born-Openheimer approximation for  atomic-molecular systems where the nuclear degrees of freedom are treated parametrically. Let us choose a coordinate system $(x, y, z)$ where the magnetic moment of the atom is written as,
\begin{equation} \label{eq:2}
\vec{\mu}=\gamma (\sigma_x \hat{i} + \sigma_y \hat{j}+\sigma_z \hat{k})
\end{equation}
where $\gamma$ is the gyromagnetic ratio. In this coordinate system, for a fixed value of parameters $x$, $y$ and $z$ the magnetic field configuration corresponding to the TOP trap \cite{three} can be expressed as ,
\begin{equation} \label{eq:3}
\vect B_{tot}= [(A_0x+B_0\cos\omega t)\hat{x} +(A_0y+B_0\sin\omega t)\hat{y} -2A_0z \hat{z}]. 
\end{equation}
We know that the field minima always lie in the $z=0$ plane and
% atom cloud is confined in the vicinity of the field minima. Due to the rotating magnetic field it rotates in $x-y$ plane (say $z=z_{min}$) 
the atom cloud also executes an oscillation \cite{ten} about the minima with the frequency $\sqrt{2}\omega_{osc}$ along the symmetry axis z. 
%((Therefore, the weak field seeker atom is necessarily confined within the specified plane)). 
%Obviously, the magnetic field and $\omega_{osc}$ depends on the the parameters $x$, $y$ and $z$. 
Eqn.\ref{eq:3} shows that the total magnetic field $B_{tot}$ of a TOP trap is rotating in  $x-y$ plane about the symmetry axis $z$. Therefore the time dependence of the Hamiltonian of the system at a fixed parametric position inside the trap appears due to the rotating magnetic field only.  Hence, the Hamiltonian corresponding to such magnetic field can be expressed as,
\begin{equation} \label{eq:4}
H = \vect \mu \cdot \vect B_{tot}.
\end{equation}  
\begin{eqnarray} \label{eq:5}
H(t)\nonumber = \gamma B(x,y,z)(\sigma_x \sin\theta(x,y,z) \cos\phi (t) \\+ \sigma_y \sin\theta(x,y,z) \sin \phi (t)+ \sigma_z \cos\theta(x,y,z)) \nonumber \\=  \omega_0(x,y,z)(\sigma_x \sin\theta(x,y,z) \cos\omega t \\ \nonumber+ \sigma_y \sin\theta(x,y,z) \sin \omega t + \sigma_z \cos\theta(x,y,z)).
\end{eqnarray}
where, $\theta$ is the angle between the z-component of the magnetic moment ($\mu$) and the magnetic field $\vec{B}_{tot}$ and $\phi(t)$ is the azimuthal angle that the magnetic field makes at the parametric point in space considered here (the value of $\theta$ can always be adjusted by choosing the suitable alignment of $\mu$ in the coordinate system, this can be tuned by sending weak field seeking state through a Stern-Gerlach apparatus). For a simpler demonstration of our idea, we have considered a two level atom whose magnetic moment is defined in terms of $2\times2$ Pauli matrices. We admit that this is not the most general case but for a multilevel bosonic or fermionic atoms, the technique can be easily generalised. We leave those complexities here. Here, we should mention that the value of the angle $\theta$  and the frequency $\omega_0$ depends on the spatial parameters $x$, $y$ and $z$ but from now on in order to investigate spin dynamics at a specific parametric point, we will write them as $\theta$ and $\omega_0$ only. The Hamiltonian in a matrix form is expressed as,
\begin{eqnarray} \label{eq:6}
\begin{aligned}
 H(t)= \frac{\omega_0}{2}\left(
\begin{array}{cc}
\cos\theta & e^{-i\omega t} \sin\theta  \\ 
e^{i\omega t} \sin\theta  & - \cos\theta \\ 
\end{array}
\right)
\end{aligned}
\end{eqnarray}   
Now, as the spin state evolves, there are two different possibilities left for the weak field seeking atom in a time dependent magnetic field. The state can either remain as weak field seeking state (adiabatic evolution) or it can flip non-adiabatically to a strong field one and flies off the trap. Obviously, the category of the evolution depends on how rapidly the Hamiltonian of the system changes with respect to the time scale characterised by the energy gap of the two level system. It is widely accepted that for a successful trapping, the system has to go through the adiabatic evolution, for which, the following criterion must be obeyed \cite{born, aharonov, amin}, 
\begin{equation} \label{eq:7}
\Big|\frac{\langle -(t)|\dot{H}| +(t)\rangle}{(E_+-E_-)}\Big|<<1. 
\end{equation}
$|+(t)\rangle$ 	and $|-(t)\rangle$ (strong and weak field seeking states) are the instantaneous eigen states of $H(t)$ and $E_+$ and $E_-$ are the corresponding energy eigenvalues. 
\[
|+(t)\rangle =
\begin{pmatrix}
e^{-i\omega t/2}\sin\theta/2\\
e^{i\omega t/2}\cos\theta/2
\end{pmatrix}
\]
and
\[
|-(t)\rangle =
\begin{pmatrix}
e^{i\omega t/2}\sin\theta/2\\
-e^{-i\omega t/2}\cos\theta/2
\end{pmatrix}
\].
 
Plugging these two states in above equation, we arrive to the condition for adiabatic evolution as,
\begin{equation} \label{eq:8} 
\frac{\omega}{2\omega_0}\sin\theta<<1.
\end{equation} 

Hence, {\it we see that $\omega<<\omega_0$, though a sufficient criteria is not necessary for the adiabatic evolution of weak field seeker state inside a TOP trap} as the angle $\theta$ also plays a crucial role. We also notice that the above ratio (Eqn.\ref{eq:8}) varies as $\theta$ and $\omega_0$ changes when the atom cloud executes its spatial motion. This gives us a clear answer to our first question we have raised in the introduction of this article. One can hope that even a weak magnetic field (for which the Larmor frequency $\omega_0$ is less than $\omega$) can be applied to trap atom by suitably choosing alignment of $\vect \mu$ in such a way that, $\sin \theta$ becomes very small. Whether this hope is at all justified or not can only be realised once we combine the spatial motion of the atom along with the spin dynamics. This issue is addressed in our article in the section IV. Obviously, the argument presented above does not apply for the parametric plane $z=0$ as $\sin \theta$ is identically one for this plane.

\section{Time evolution of a weak field seeking state at a given parametric position inside TOP trap}

At this juncture, $\*$let's turn back to the second important question raised earlier in this article. To answer this, one must solve Schr\"odinger equation for the Hamiltonian in Eqn.\ref{eq:4} with the initial state $|-\rangle$ at $t=0$. Obviously by doing so, we are able to know the dynamics of spin state by keeping the atom at some fixed position. The evolved state can be written as the linear combination of the instantaneous eigenstates of time dependent Hamiltonian $H(t)$. Let's call it, 
\begin{equation} \label{eq:9}
|t\rangle = \alpha(t)|-\rangle_t + \beta(t)|+\rangle_t.
\end{equation}
Added to that, we avoid any time averaging concept and hence by making it a general treatment applicable to any arbitrarily small time scale.\\

\begin{figure}
\begin{minipage}{.4\textwidth}
\includegraphics[width=\textwidth]{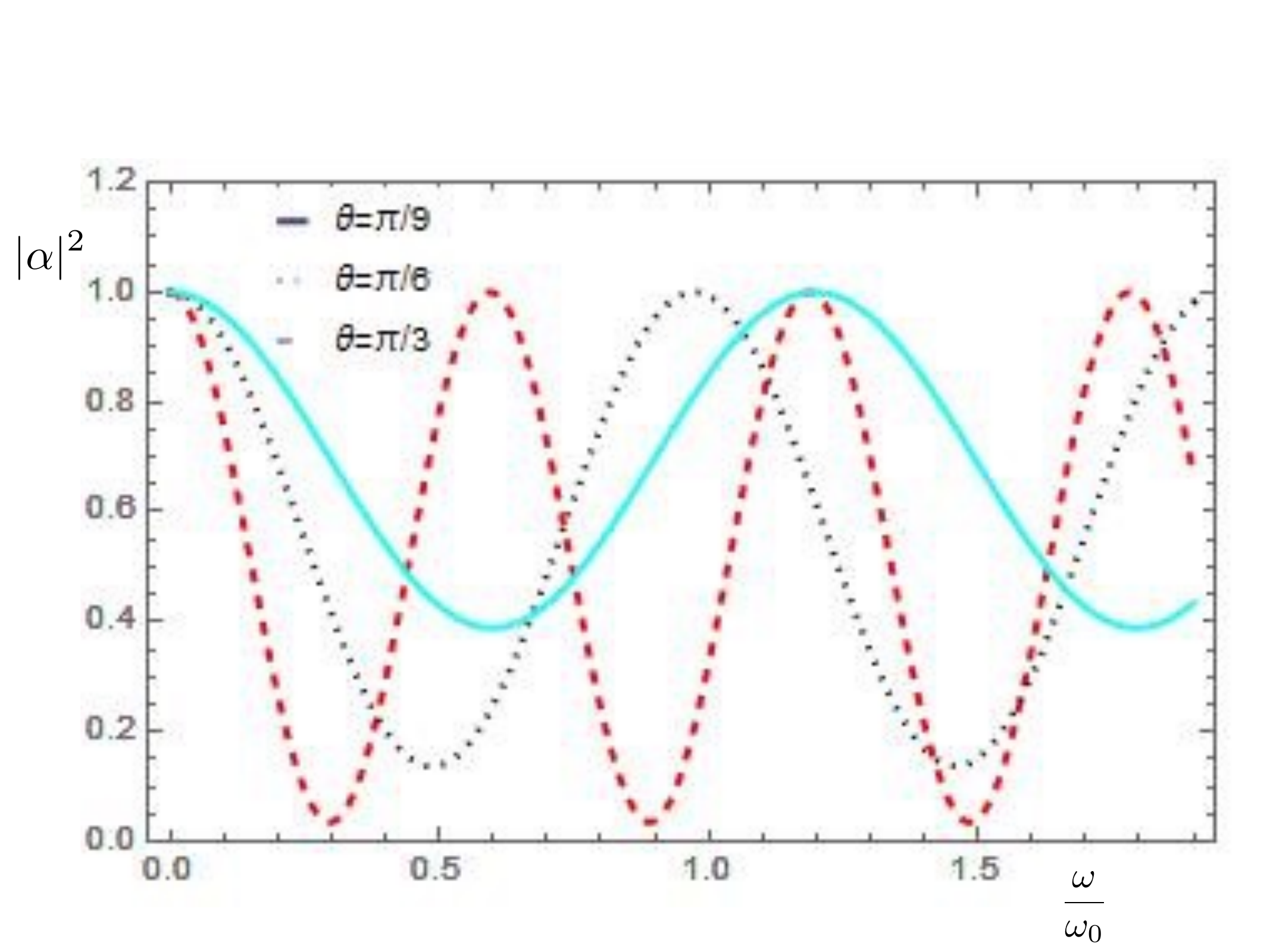}
\caption{\small The oscillatory survival probabilities of weak field seekers for a given value of $\omega= 1.5\omega_0$, show different amplitudes of oscillation for different $\theta$.\\ }
\label{fig:one}
\end{minipage}

\begin{minipage}{.4\textwidth}
\includegraphics[width=\textwidth]{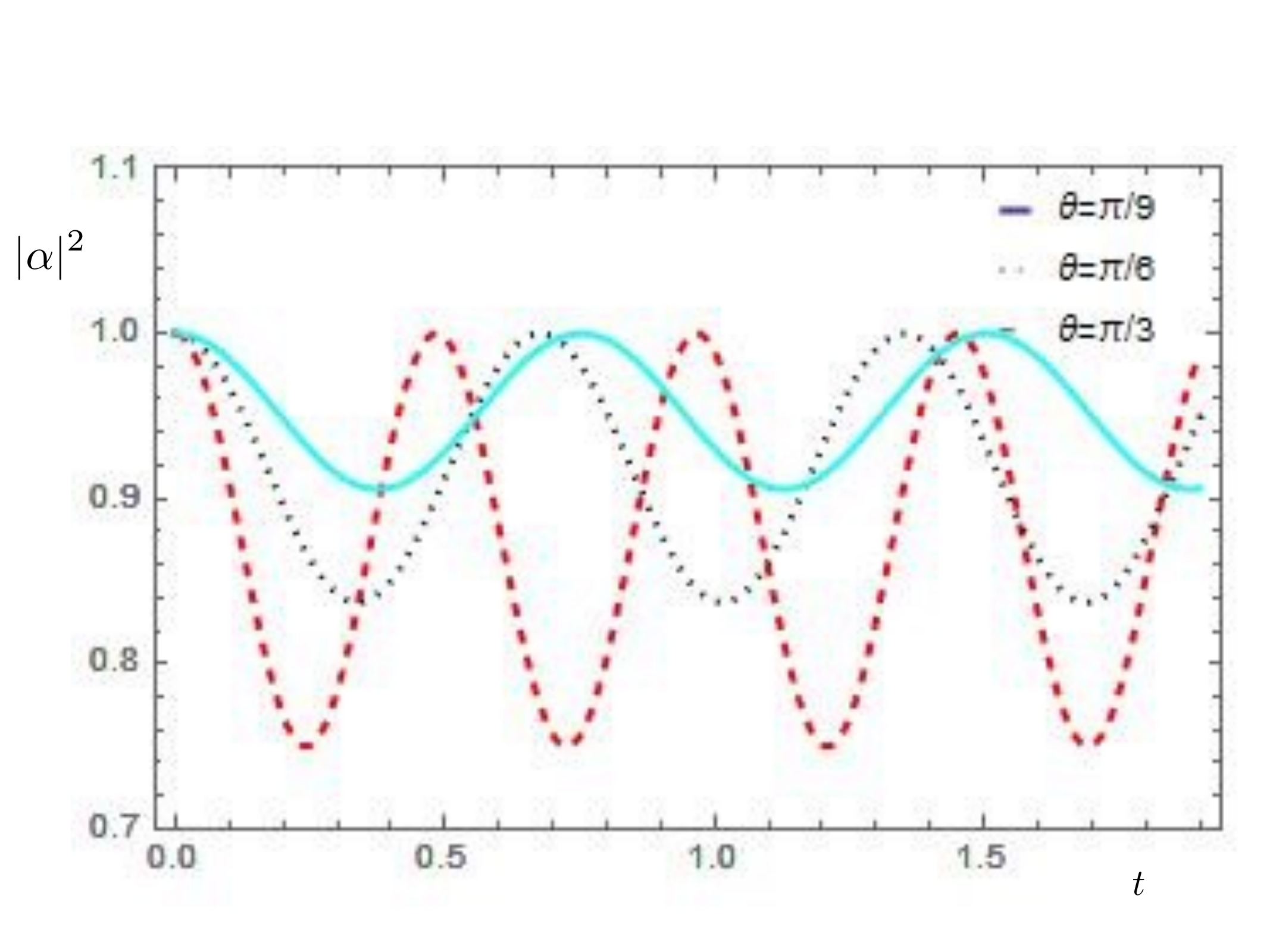}
\caption{\small For a comparatively lesser value of $\omega(\omega=0.5\omega_0)$, the survival probabilities for different values of $\theta$ show comparatively smaller values of amplitudes of oscillation.\\}
\label{fig:two}
\end{minipage}
\end{figure}

Plugging Eq.(7) in the Schr\"odinger equation, 
\begin{equation}  \label{eq:10}
i\frac{d}{dt} (\alpha(t)|-\rangle_t + \beta(t)|+\rangle_t) = H(t)|t\rangle,
\end{equation}  
we arrive to a set of coupled differential equations \cite{twelve, thirteen}, 
\begin{equation}  \label{eq:11}
\dot{\alpha}(t) + \frac{i}{2}\left[\omega_0 - \dot{\phi}(t)\cos\theta\right]\alpha(t)-\frac{i}{2}\dot{\phi}(t)\beta(t)\sin\theta = 0
\end{equation}
\begin{equation}  \label{eq:12}
\dot{\beta}(t) - \frac{i}{2}\left[\omega_0 - \dot{\phi}(t)\cos\theta\right]\alpha(t)-\frac{i}{2}\dot{\phi}(t)\alpha(t)\sin\theta = 0
\end{equation}
Hence, by solving them with the initial condition $\alpha(0)= 1$ $\&$ $\beta(0) =0$, we have 
 \begin{equation}  \label{eq:13}
\alpha(t)= \cos \frac{\bar{\omega} t}{2} + \frac{i}{2}\left[\omega_0- \omega \cos\theta\right]\sin \frac{\bar{\omega}t}{2} 
\end{equation}
\begin{equation}  \label{eq:14}
\beta(t) = \frac{i\omega \sin\theta}{\bar{\omega}} \sin\frac{\bar{\omega}t}{2}
\end{equation}
Where,  $\bar{\omega} = \sqrt{\omega_0^2+\omega^2-2\omega_0 \omega \cos\theta}$\\
From these two coefficients $\alpha(t)$ $\&$ $\beta(t)$ we can calculate the survival probability of weak field seeking state at any instant t, given by,
\begin{equation}  \label{eq:15}
|\alpha(t)|^2 = \cos^2 \frac{\bar{\omega}t}{2}+\frac{\left[\omega_0-\omega \cos\theta\right]^2}{\bar{\omega}^2} \sin^2\frac{\bar{\omega}t}{2}.
\end{equation}
Similarly, the transition probability (loss) from weak field to strong field seeking state is given by,
\begin{equation}  \label{eq:16}
|\beta(t)|^2 = \frac{\omega^2 \sin^2\theta}{\bar{\omega}^2}\sin^2\frac{\bar{\omega}t}{2}.
\end{equation} 
Now, we consider a case where $\omega_0$ is not much larger than $\omega$ and $\theta$ is not negligibly small. In that case the time evolution of weak field seeking state is absolutely non adiabatic. It is evident from above equations that the survival and transition probability vary with the position of the atom cloud inside the trap as $\theta$ and $\omega_0$ is a function of $x$, $y$ and $z$. The atom cloud moves inside the trap by executing oscillations along all the coordinate axes. % For very small radius of circle of death of the atom cloud, one can consider that the atom cloud executes an oscillation along $z$ and the goes through different parametric planes ($z$=constant) while executing oscillation about $z=0$. 
Therefore, it is customary to show the nature of survival probability at different parametric position of the atom characterised by different value of $\theta$ and $\omega_0$. The survival probability $|\alpha(t)|^2$ shows an oscillatory pattern which is shown in Fig.\ref{fig:one} and Fig.\ref{fig:two}, considering two different values of $\omega$ $(\omega= 0.5\omega_0,  \omega = 1.5\omega_0)$ for different $\theta$. It is evident that the amplitude of oscillations decays with both $\theta $ and $\frac{\omega}{\omega0}$ and becomes constant (equals to unity) for $\omega_0>>\omega$ or  for $\theta = 0$, which eventually boils down to the adiabatic evolution of weak field seeking state.   

\begin{figure}
\begin{minipage}{.4\textwidth}
\includegraphics[width=\textwidth]{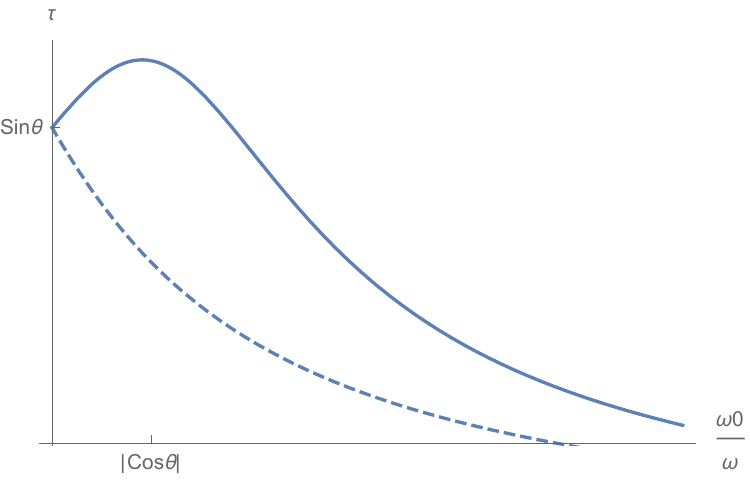}
\caption{\small The solid line shows that for the parameter  region $0<\theta\leq \frac{\pi}{2}$, the resurrection time $\tau$ reaches a local maxima at $\frac{\omega0}{\omega}=|\cos \theta|$ and then decreases with $\frac{\omega0}{\omega}$ and  the dashed line shows that $\tau$ decreases monotonically with $\frac{\omega0}{\omega}$ for the parameter region $\frac{\pi}{2}\leq\theta \leq\pi$.}
\label{fig:three}
\end{minipage}
\end{figure}

This is indeed very interesting as we can see that, though the weak field seeker state flips into a strong field seeking one, resurrects again after a time $\tau_{res}=\frac{2\pi}{\bar{\omega}}$. In the unit of $\frac{2 \pi}{\omega}$, $\tau_{res} $ can be expressed as,
\begin{equation}  \label{eq:17}
\tau = \frac{1}{\sqrt{1+(\frac{\omega_0}{\omega})^2-2(\frac{\omega_0}{\omega})\cos\theta}}.
\end{equation}

The resurrection time $\tau$ is an emergent time scale, composed of $\omega$, $\omega_0$ and the parameter $\theta$. Now, the trap for which the weak field seekers take more than the resurrection time to fly off, are permanently confined because, before leaving the trap, they become weak field seekers again, and hence, are pulled towards the minima and it is always possible to design such a trap accordingly. So, upto this point, {\it our study offers a tentative realisation that atoms might not fly off the trap even though they have gone through a non adiabatic evolution.} The complete answer is possible only when we will combine the spatial motion along with the evolution of the spin state. We will present such a calculation in the next section of this article which will provide the answer to our second query. \\
For more precise and quantitative understanding of $\tau$, we have presented two plots corresponding to two different regimes of the parameter $\theta$. Fig.\ref{fig:three} shows that within the range $0\leq\theta\leq\frac{\pi}{2}$ the quantity $\tau$ initially increases from unity and reaches a local maxima equals to $\frac{1}{\sin \theta}$ at ($\frac{\omega_0}{\omega}=|\cos \theta|$) and then decreases with ($\frac{\omega_0}{\omega}$). Therefore, for a fixed value of $\theta$ within the above mentioned range, $\tau$ never can be bigger than cosec$\theta$. Similarly, for $\pi/2<\theta\leq\pi$ , $\tau$ monotonically decreases with ($\frac{\omega_0}{\omega}$). For both the above mentioned domains of $\theta$, a very large value of ($\frac{\omega_0}{\omega}$) corresponds to the adiabatic domain of time evolution, where $\tau$ is almost equals to zero. This simply signifies that, the weak field seeker states never flip to strong field seeker states during the course of their evolution. On the other hand, when $\frac{\omega_0}{\omega}$ is infinitesimally small the quantity $\tau$ becomes unity. Now, for a very large value of $\omega$, a weak field seeker atom takes infinitesimally small time to resurrect completely. This is extremely counter-intuitive from the conventional understanding of TOP trap, as in such a situation the evolution of atoms is highly non adiabatic only with exceptions when $\theta$ is either close to zero or $\pi$. In other words, a highly non adiabatic case might be a favourable situation for a successful trapping of atoms. 

\section{Average survival probability of weak field seeking state inside a TOP trap}
\begin{figure}
\begin{minipage}{.4\textwidth}
\includegraphics[width=\textwidth]{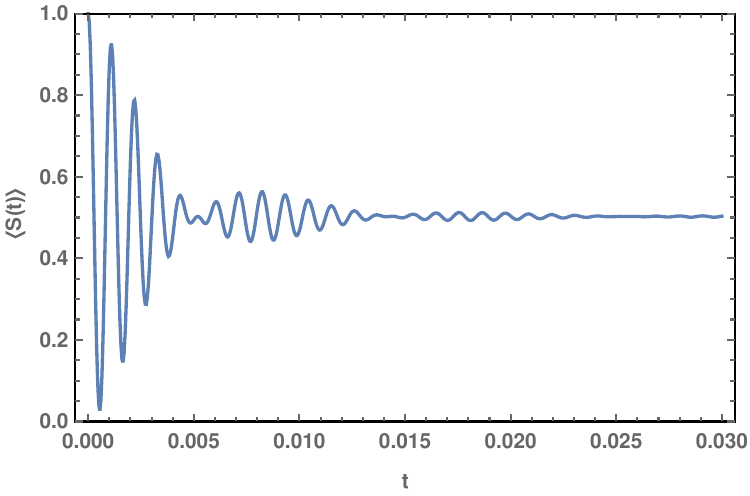}
\caption{\small The plot shows how the spatial average of survival probability of weak field seeking atom changes with time for a chosen set of free parameters $A_0 = 110 G/cm, B_0 = 10^{-1} G$ and $\omega = 20\times 10^5 Hz$.}
\label{fig:four}
\end{minipage}
\end{figure}

\begin{figure}
\begin{minipage}{.4\textwidth}
\includegraphics[width=\textwidth]{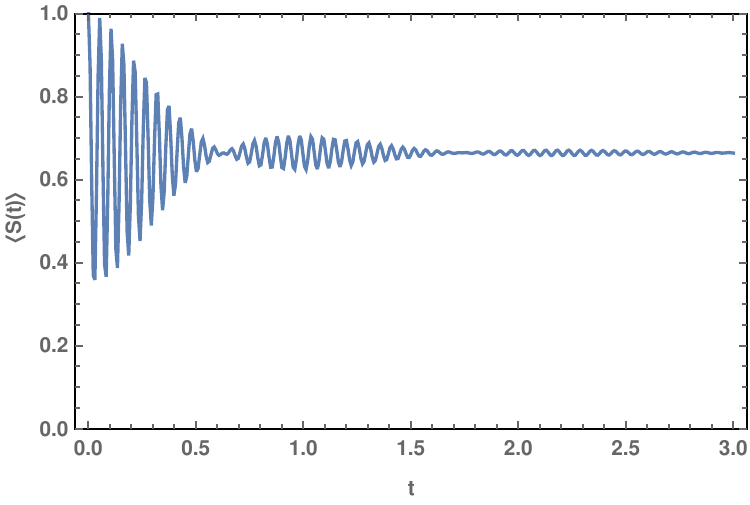}
\caption{\small The plot shows how the spatial average of survival probability of weak field seeking atom changes with time for a chosen set of free parameters  $A_0 = 110 G/cm, B_0 = 10^{-2} G$ and $\omega = 20\times 10^4 Hz.$.}
\label{fig:five}
\end{minipage}
\end{figure}

In the last two sections, we have discussed dynamics of spin states considering the atom cloud at a parametric point in space. The atom cloud passes through various spatial points while executing oscillation along $x$,  $y$  axes with a frequency $\omega_{osc}$ and along $z$ axis with a frequency $\sqrt{2}\omega_{osc}$. The survival probability varies from one parametric space point to another. Therefore, the meaningful quantity one can calculate is the  spatial average of survival probability of the atom inside the TOP trap. For that we must have to consider the spatial motion of the atom accordingly. The average survival probability can be expressed as 
\begin{equation}  \label{eq:18}
<S> = \int dx dy dz |\psi_{-}(x, y, z, t)|^2 |\alpha(x, y, z, t)|^2,
\end{equation} 
where $\psi_{-}(x, y, z, t)$ is the spatial ground state wave function of the atom at some instant $t$. The wave function can be evaluated by solving the two component Pauli equation for the atom 
\cite{six}, it looks 

\begin{eqnarray}  \label{eq:19}
\psi_{-}(x,y,z,t) = \left(\frac{m\omega_{osc}}{\pi\hbar}\right)^{\frac{1}{4}}exp\left[\frac{-m\omega_{osc} x^2}{2\hbar} \right]\nonumber  \\ \nonumber \left(\frac{m\omega_{osc}}{\pi\hbar}\right)^{\frac{1}{4}} exp\left[\frac{-m\omega_{osc} y^2}{2\hbar} \right]\left(\frac{m\omega_z}{\pi\hbar}\right)^{\frac{1}{4}}exp\left[\frac{-m\omega_z z^2}{2\hbar} \right] \\ \nonumber exp\left[\left(\frac{if}{\hbar\omega}\right)(x\sin\omega t - y\cos\omega t)-i(\omega_l+E_{nlq}/\hbar)t)\right],\\ 
\end{eqnarray}
 where $\omega_l = |\mu|B_0/\hbar$ is the Larmor frequency at the centre of the trap, $\omega_z = 2 \sqrt{2}  \omega_{osc}$, $f=|\mu|A_0$ is the magnitude of the force rotating in the $x-y$ plane with frequency $\omega$ and $E_{nlq}$ is the eigenenergies corresponding to the time-independent eigenfunctions\cite{six}. \\
 Now, in principle, one can calculate the average survival probability of weak field seeking atoms inside the trap using Eqn.\ref{eq:18} for different TOP trap parameters such as quadrupole magnetic field gradient $A_0$, bias magnetic field $B_0$ and frequency $\omega$ of the rotation of the bias field. We have chosen two different sets of free parameters such as, $A_0 = 110 G/cm, B_0 = 10^{-1} G$ and $\omega = 20\times 10^5 Hz$ and another set for $A_0 = 110 G/cm, B_0 = 10^{-2} G$ and $\omega = 20\times 10^4 Hz$ for which the average survival probability as function of time is presented in the Fig.\ref{fig:four} and Fig.\ref{fig:five}. For these sets of parameters the Larmor frequency, $\omega_l= 1.4\times 10^{5} Hz$ and $\omega_l= 1.4\times 10^{4} Hz$ respectively, which is less than the frequency of the rotating bias field $\omega$. Not only at the centre of the trap, it can be checked that for these sets of top trap parameters, adiabatic condition has been violated in some other region of the trap also but still the weak field seeking state can be confined inside the trap with certain amount of loss. Fig.\ref{fig:four} and Fig.\ref{fig:five} show that the average survival probability of atom initially oscillates with time and stabilises after sometime.  For first set of parameters the stabilised value of average survival probability is around $0.5$ and the stabilisation time is 30 milliseconds whereas for the second set of parameters these two quantities change to $0.67$ and 3 seconds respectively. These results nicely demonstrate the answers of the two queries which we have been raised in the beginning of the article.

\section{Conclusions}
To conclude, we have explored two important aspects of a TOP trap. First one is the necessary and sufficient condition for adiabatic evolution of weak field seeker states inside the trap. In this context, for the first time, we have realised the importance of the parameter $\theta$, which plays a crucial role in the determination of the category of time evolution for a given parametric position inside the trap. For a suitable value of the parameter $\theta$, we can restore the adiabatic evolution of weak field seeking states no matter how large is the frequency $\omega$ compared to the Larmor frequency ($\omega_0$) of the system. This gives us a freedom to relax the well known criterion (1) for successful trapping of atoms. Added to that we have pointed out that the atom can also be trapped instead of their non adiabatic evolution. We have arrived to these two conclusions with a purely quantum mechanical approach without adopting any time averaged description of the magnetic field. This also helps us avoid semi classical  hitherto conventional treatment on the specific topic. On the experimental side, our findings offers a nice opportunity for a less restricted magnetic trap.
Now, even employing a weak bias field one can successfully trap atoms in the laboratory. Concluding our study we must say {\it though the adiabatic criterion is employed as a condition for successful trapping of atoms over the years but not necessary at all}. We have realised that we can trap atoms even in the non-adiabatic regime and we hope this can be very useful and cost effective approach in the cold atom laboratory. We admit that so far our analysis is based on the consideration of a two level system but a technical generalisation for multilevel system is no longer far removed. We leave that task aside for a separate article in near future.\\

%%%%%%%%%%%%%%%%%%%%%%%%%%%%%%%%%%%%%%%%%%%%%%%%%%%%%%%%%%%%%%%%

\begin{acknowledgements}
We acknowledge Prof. Debasish Mukherjee, Prof. W. Ketterle and Prof. W. V. Klitzing for insightful discussions. A. D. acknowledges IISER Kolkata for supporting the project and Rafiqul Rahaman , Sreeraj Nair for fruitful discussions. 
\end{acknowledgements}

\bibliography{ref}{}
\bibliographystyle{unsrt}

\end{document}